\documentclass[aps,showpacs,amsmath,amssymb,nofootinbib,preprintnumbers]{revtex4}

\usepackage{graphicx}
\usepackage{color}

\def \be  {\begin{equation}}
\def \ee  {\end{equation}}
\def \bea {\begin{eqnarray}}
\def \eea {\end{eqnarray}}

\begin{document}

\preprint{ECTP-2010-04}

\title{Acceleration and Particle Field Interactions of Cosmic Rays I: Formalism}

\author{A.~Tawfik}
\email{drtawfik@mti.edu.eg}
\affiliation{Egyptian Center for Theoretical Physics (ECTP), MTI University,
 Cairo-Egypt}
 \author{A.~Saleh}
\affiliation{Egyptian Center for Theoretical Physics (ECTP), MTI University,
 Cairo-Egypt}
 \author{M.~T. Ghoneim}
\affiliation{Physics Department, Faculty of Science, Cairo University, Cairo-Egypt}
 \author{A.~A. Hady}
\affiliation{Astrophysics Department, Faculty of Science, Cairo University, Cairo-Egypt}


\begin{abstract}

The acceleration of ultra high energy cosmic rays is conjectured to occur through various interactions with the electromagnetic fields in different astrophysical objects, like magnetic matter clumps, besides the well-known shock and stochastic Fermi mechanisms. It is apparent that the latter are not depending on the particle's charge, quantitatively. Based on this model, a considerable portion of the dynamics, that derives a charged particle parallel to a magnetic field $\mathbf{B}$ and under the influence of a force $\mathbf{F}$, is assumed to be composed of an acceleration by a non-magnetic force $\mathbf{F}_{\parallel}$ and a gyromotion along $\mathbf{B}$ direction, plus drifts in the direction of $\mathbf{F}_{\perp}$. The model and its formalism are introduced. Various examples for drift motions and accelerating forces are suggested. The formalism is given in a non-relativistic version. Obviously, the translation into the  relativistic version is standard. In a forthcoming work, a quantitative estimation of the energy gained by charged cosmic rays in various astrophysical objects will be evaluated. 

\end{abstract}

\pacs{96.50.S-, 03.50.De, 41.20.-q, 52.35.Kt}


\maketitle

\section{Introduction}

One of the greatest mysteries in modern physics is the origin of ultra high energy cosmic rays (UHECR). The current acceleration technologies are unable to come up with such energy scales. In order to compare, let us mention the several observatories in which cosmic rays with energies $>10^{20}~$eV have been registered~\cite{pap0,pap1,pap2,pap3,pap4}. From the theoretical point of view, the protons, which are the dominant components of UHECRs, would posses energy values much higher than the so-called Greisen-Zatsepiti-Kuzmin (GZK) cutoff $10^{19.7}~$eV. They are conjectured to suddenly lose their energies in photopion production processes with the cosmic microwave background (CMB) radiation \cite{pap5,pap6,pap7}. This model is known as the top--down scenario. 

In this work, we study the acceleration of cosmic charged particles starting from low energies, i.e. bottom-up scenario. The kinematics of such particles that move parallel to a magnetic field $\mathbf{B}$ is assumed to be composed of an acceleration through a non-magnetic force $\mathbf{F}_{\parallel}$ and a gyromotion along the direction of $\mathbf{B}$ field plus the drift(s) in the direction of $\mathbf{B}\times\mathbf{F}_{\perp}$, as illustrated in Fig. \ref{fig:1a}. The drift motions are not depending on the Larmor radius and therefore can exist even in cold plasmas. This would partly explain why a non-relativistic formalism is explicitly given here. The drift motions are the summation of all types of the perpendicular velocities stemming from different non-magnetic forces. On the other hand, the stochastic acceleration according to the second-order Fermi mechanism apparently stems from the accumulation of a small velocity change with non-negligible random velocity components in each elastic collision with the magnetic cloud. The magnetic cloud can be considered as a clump of a homogeneously distributed matter moving with Alfven wave velocity $v_A=B_0/\sqrt{4\pi\,\rho}\,$, where $B_0$ and $\rho$ are strength of {\it background} magnetic field and {\it background} matter density, respectively. With {\it background} we mean the cloud's {\it local} frame of reference. The final acceleration occurs as a consequence of the field interactions and multiple scatterings of the charged particles with the electromagnetic irregularities. 

To answer the question about the possibility of finding both magnetic and electric fields in various astrophysical objects, let us first remember that the magnetic fields are basic characters of the cosmic matter clumps. The existence of electric fields would be explained by the cloud's plasma properties \cite{parks}. The active galactic nuclei (AGN) and gamma-ray bursts (GRB) are often considered as plausible sources for the astrophysical accelerators of extragalactic UHECR \cite{agn-grb1}. The highly magnetized neutron stars    \cite{nstars} and the structure formation shocks \cite{shocks} are conjectured as possible sources. Also, the spinning black holes have been suggested \cite{BH-accel} to accelerate the cosmic charged particle. The existence of electric fields in all these astrophysical objects is obvious. On one hand, they would sanctify that the cosmic rays likely participate in various field interactions. On the other hand, they might be baptized as Eevatrons. They would provide energy scales up to several hundred EeV ($10^{21}\,$eV). 

In 1984, Hillas introduced a widely celebrated approach \cite{hillas1}. Magnetic field interactions are assumed to accelerate the cosmic charged particles. In present work, we suggest an extension to this pioneering work. Both electric and stochastic interactions are also taken into consideration. Furthermore, we assume a general configuration for the electromagnetic fields. We are not exclusively considering the very special configuration that magnetic and electric forces are perpendicular to each other. The latter apparently limits the gyration to a concrete minimum value. It is conjectured that various interaction types might be possible in one astrophysical object. The final energy is then accumulated over successive acceleration steps. 

\begin{figure}[bht]
\includegraphics[height=3.5cm]{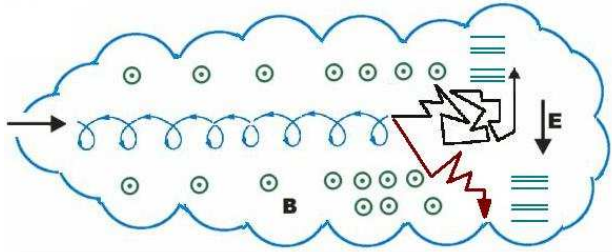}
\caption{\footnotesize An illustration of the model applied in this letter with {\it ideal} configuration. The magnetic cloud embeds regions of static and spatiotemporal-varying electromagnetic fields, so that the charged particles likely experience both field interactions and elastic scattering. } 
\label{fig:1}
\end{figure}

In present work, both first- and second-order Fermi mechanisms remain playing their essential roles. It is worthwhile to mention that the energy gained according to Fermi mechanisms explicitly depends on the relative particle's velocity $v$ with respect to the velocity of the rest frame $u$ of the magnetic clouds and/or matter clumps. Various types of field interactions have been added to the stochastic interactions of Fermi. This reflects among others the quantum properties of the cosmic particles. An illustration of this approach is given in Fig. \ref{fig:1}. 

In general, it turns to be no longer a trivial task to suggest a new mechanism in order to explain the acceleration of  cosmic charged particles to such ultra high energy. Many constrains have to be taken into account. One of these constrains would be the requirement that the particle must be imprisoned inside the range of the active field while it is being accelerated. Therefore, one has to be aware with the particle gyroradius. Other constrains would include the strengths of accelerating fields and the time interval till the particle reaches EeV-energy-scale. Also, one has to take into consideration the various fluctuations and the possibilities for energy loss. In present work, we highlight the properties of cosmic particles \cite{cr_comps}. They are electrically charged and eventually carry other quantum numbers, like mass and angular momentum. Over their long path to the Earth, they go through various plasma fields and elastic/stochastic collisions. Therefore, various types of electromagnetic field interactions have to be added to Fermi acceleration mechanisms in order to compute the final energy.  

\section{Model}
\label{sec::1}

In present work, we suggest a model, which fulfills many of the constrains discussed in the previous section. Since electric and magnetic fields can also exist in cold plasmas, we explicitly introduce a non-relativistic formalism. The translation into a relativistic version is straightforward. We therefore leave it to an upcoming work, which will be devoted to the quantitative calculation of the energy gained by cosmic charged particles. 

\begin{figure}[thb]
\includegraphics[height=3.5cm]{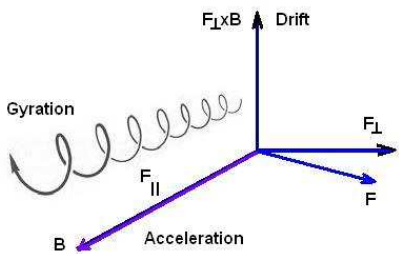}
\caption{\footnotesize Trajectory of motion of a charged particle in a non-uniform magnetic field under the influence of a non-magnetic force $\mathbf{F}$. } 
\label{fig:1a}
\end{figure}

Let us consider a charged particle with mass $m$ and static electric charge $q$ in an electromagnetic field characterized be $\mathbf{E}$ and $\mathbf{B}$. $\mathbf{E}$ and $\mathbf{B}$ being globally conserved electric and magnetic fields, respectively. This assumption does not prevent that both fields can be varying, at least locally. We also assume that particle undergoes the well-known elastic , stochastic \cite{fermi49} and shock \cite{shockB} Fermi mechanisms. In addition to the stochastic scattering, which mainly changes the particle's initial velocity $v$ by $\pm 2u$, we assume that the particle simultaneously interacts with the magnetic and non-magnetic forces, Fig. \ref{fig:1a}. These forces are not necessarily uniform or constant.

First, we like to refer to the fact that magnetic field in electromagnetic interactions is not necessarily participating in the acceleration mechanism. Accordingly, it merely rotates or redirects the particle's motion. This is apparently restricted to the so-called ''$\mathbf{E}\times \mathbf{B}$'' drift velocity in uniform electromagnetic fields. In this case, the electric and magnetic forces are simply given by $q\mathbf{E}$ and $q \mathbf{v}\times\mathbf{B}$, respectively. The guiding center is given as $\mathbf{v}_{\mathbf{E}}=(\mathbf{E}\times\mathbf{B})/B^2$. The subscript $\mathbf{E}$ in this case refers to the fact that $\mathbf{E}$ derives a drift velocity independent on the background properties but referring to the interactions with $\mathbf{B}$. The uniform configuration of the electromagnetic fields would limit the discussion to very special astrophysical phenomenon. But the galactic dipolar fields and loops, for instance, are poorly described by this special kind of the configurations.

In present work, we study general configurations of electromagnetic fields. They are not necessarily perpendicular to each other. Furthermore, we take into consideration that they are spatially and temporally varying. The effects of gravitational fields should not be entirely excluded, especially, when the gravitational forces of extremely gigantic astrophysical objects are dominant \cite{BH-accel}.

\section{Formalism}

\subsection{Cosmic charged particle resting in the magnetic cloud}

We first assume that the cosmic charged particle in the cloud's frame of reference fulfills two conditions; resting at a certain point, where the electromagnetic potential is nearly vanishing and entering the cloud with an initial velocity. These two cases are graphically illustrated in Fig. \ref{fig:1}. It is conjectured that the forces affecting the particle are based on Lorentz ($\mathbf{E}$,$\mathbf{B}$) and gravitational $\mathbf{{\cal G}}$ fields. In natural units, the equation of motion of a charged particle reads
\bea
\label{lorentz1}
m\frac{d\,v}{d\,t} = q \left[\mathbf{E}(\mathbf{r},t) + \mathbf{v}\, \times\, \mathbf{B}(\mathbf{r},t)\right] + m \mathbf{{\cal G}}(\mathbf{r},t)
\eea
where $\mathbf{{\cal G}}$ is the effective gravitational acceleration. For simplicity, we assume that $\mathbf{r}$ gives a certain one-dimensional direction that is parallel to the lines of $\mathbf{B}$--field. 
According to the assumption, which is frequently applied in literature, that the gravitational fields are much weaker than the electromagnetic ones, especially for light particles, then the particle moves with the velocity
\bea  \label{eq:ebv}
\mathbf{v} &=& \sqrt{-\frac{2\, q\, \mathbf{\phi}}{m}},
\eea 
and acceleration 
\bea \label{eq:eba}
\mathbf{a} &=& \frac{1}{\sqrt{-2\, q\, m\, \mathbf{\phi}}} \left(\mathbf{\phi} \mathbf{I} + q \frac{d\mathbf{\phi}}{d t} \right) = \sqrt{\frac{2\, q\, \mathbf{E}}{m\, r}} \; v,
\eea
where $\mathbf{\phi}$ and $\mathbf{I}$ being the electric potential and current along $\mathbf{r}$, respectively. Obviously, the electromagnetic fields -- in these expressions -- are conjectured to be static and uniform. In Eq. (\ref{eq:eba}) and seeking for simplicity, we assume that $\mathbf{E}$ makes an angle $\theta=\pi/4$ with $\mathbf{B}$. Furthermore, we assume that both $\mathbf{v}$ and $\mathbf{B}$ are parallel, i.e. in magnetohydrodynamics (MHD) $\mathbf{E}_{\perp}=\mathbf{J}/\sigma-\mathbf{V}\times \mathbf{B}$, where $\mathbf{V}$ is the velocity of the background plasma. $\mathbf{J}$ is the current density and $\sigma$ is the charge conductivity. Then, the cosmic charged particle would gain a nett acceleration resulting from the parallel acceleration and drifts across $\mathbf{B}$--field. Its direction is given by the parity image of $\mathbf{E}$ on a mirror allocated along the lines of $\mathbf{B}$--field. The same is valid for the particle's guiding center.
\bea \label{eq:ebenrg}
\mathbf{a} &=& \sqrt{\frac{\sqrt{2}\, q}{m\, r}\, \left(\mathbf{E}_{\parallel} - \mathbf{V}\times \mathbf{B}\right)} \; v,
\eea
where an ideal MHD has been assumed. $\mathbf{r}$ is the distance covered by the particle, which can be related to the Larmor radius $\mathbf{r}_L$ \cite{larmor}. $\mathbf{r}_L$ is defined as the radius of the cyclotron motion derived by the magnetic force. The motion of the changed particles under the influence of these static electromagnetic fields can be decomposed into a Larmor motion in the absence of electric fields and a drift of gyration center of its rotation. This would mean that the total velocity is composed of three components $\mathbf{v}_{\perp}$, $\mathbf{v}_{\parallel}$ and $\mathbf{v}_D$. Obviously, the maximum acceleration can be obtained when $\mathbf{E}$ and $\mathbf{B}$ are parallel and the charged particle -- as usual -- moves along the lines of $\mathbf{B}$--field. 

The energy that the cosmic charged particle would gain through this field interaction reads
\bea\label{eq:eF2_NoFermi}
\epsilon &=& \sqrt{\sqrt{2} \, m\,r \, q \left(\mathbf{E}_{\parallel} - \mathbf{V}\times \mathbf{B}\right)} \; v,
\eea
where Fermi acceleration mechanism has been excluded. It is obvious that the final motion is a nett result out from two forces. The first one is the parallel component of the electric field strength, $\mathbf{E}_{\parallel}$. The second force is the drift from the perpendicular component.  

A general expression for the drift motion would inhere in it modifications of the inhomogeneity (non-uniformity) of the electric field $\mathbf{E}$.  
\bea\label{eq:ve}
\mathbf{v}_\mathbf{E}&=&\left(1+\frac{r^2_L\nabla^2}{4}\right) \frac{\mathbf{E}\times\mathbf{B}}{B^2},
\eea 
where $\mathbf{E}=\mathbf{E}_{\parallel}\cos(\theta)+\mathbf{E}_{\perp}\sin(\theta)$ can be replaced by arbitrary constant force. The physics remains unchanged. For example, the gravitational drift velocity $v_{\mathbf{a}_g}=(m/qB^2)\mathbf{a}_g\times\mathbf{B}$, where $\mathbf{a}_g$ is the gravitational acceleration in the magnetic cloud. The dependence on mass $m$ and charge $q$ implies that this drift can be neglected for light particles and its direction changes with the sign of the electric charge $q$. The gravitational force is proportional to the product of masses of the two interacting counterparts. Therefore, it is likely dominant, when one of the counterparts would be -- for instance -- a black hole. The mass of black holes would range from $m_{BH} \approx(0.5-1.2)\times 10^8\,M_{\odot}$ \cite{bh-mass}, where $M_{\odot}$ is the solar mass.

We notice that Eq. (\ref{eq:ve}) seems to reflect a special case, in which the electric field $\mathbf{E}$ is spatially non-uniform and oriented perpendicular to the plane of $\mathbf{z}$ and $\mathbf{y}$. As given in Fig. \ref{fig:1}, $\mathbf{z}$ is the direction of $\mathbf{B}$--field. This equation would mean that the average value of drift motion of a charged particle that apparently spends more time in regions with weak $\mathbf{E}$ field strength is less than $\mathbf{E}\times\mathbf{B}$ value. The latter is to be computed at the guiding center (the center of gyration of the particle of interest) \cite{somov}. 

Within one Larmor orbit, the electromagnetic fields are assumed to be uniform. In other words, the characterizing length over which the fields vary is much smaller than $\mathbf{r}_L$. In this case, the gyration orbit can be approximated to a closed circle. The differential drift would result in a net current density, $\mathbf{I}=\sum_r\, n_r\, q_r\, \mathbf{v}_{\mathbf{E},r}$. 

It is obvious that the field interactions are positioned within the boundaries of the magnetic cloud and therefore {\bf $r_L\lesssim r$} \cite{hillas1}. Then 
\bea \label{rrl_special}
\mathbf{r}_L&=&\frac{m}{q\,B}\;\mathbf{v}=\frac{1}{\pm\,\Omega\,t}\,\mathbf{r},
\eea
where $\Omega$ is the angular velocity. $\pm$ stands for right- and left-handed rotation of the Larmor motion. These signs are assumed to be given by the electric charge $q$. $v_{\perp}$ is the velocity component which is oriented  perpendicular to the $x$--$y$ plane. The relation between $v_{\perp}$ and the scalar $v_L$ is given by $v_L=\sqrt{\mathbf{v}_{\perp} \cdot \mathbf{v}_{\perp}}$. At an arbitrary angle $\Omega\,t$, expression (\ref{rrl_special}) would read
\bea\label{rrl_general}
\mathbf{r}_L &=& \frac{\mathbf{v}}{\Omega} \left(\hat{x}\sin \Omega t + \hat{y}\cos \Omega t\right)
\eea
Alternatively, we can substitute Eq. (\ref{eq:ebv}) into Eq. (\ref{eq:eba}), then the energy that the charged particle would gain through the electric field $\mathbf{E}$ is given as
\bea \label{energ12}
\epsilon &=& 2 \, q\, \mathbf{r}\cdot\mathbf{E}
\eea

In deriving above expressions, we assume that the electric field strength $\mathbf{E}$ is responsible for the releasing of a charged particle from rest and then accelerating it to the energy given in Eq. (\ref{energ12}). Equation (\ref{energ12}) represents the basic of the well--celebrated Hillas graph \cite{hillas1}. It is worthwhile to note that this is a special case. This has been discussed above. The general case would require further treatments.

\subsection{Cosmic charged particle entering the magnetic cloud}

When the charged particle is assumed to enter the magnetic cloud - or in general regions of magnetic irregularities -, then the initial conditions can be very much different. At a spacial distance $\mathbf{r}_i$, the initial kinetic energy would take the form $m \mathbf{v}_i^2/2$. Then, Eq. (\ref{lorentz1}) leads to the following equation of motion
\bea
\mathbf{v}_f^2 - \mathbf{v}_i^2 &=& \frac{2}{m} q \int_{r_i}^{r_f} \mathbf{E} \cdot d\mathbf{r},
\eea
which apparently depends on $\mathbf{E}$, exclusively. Since, 
both electric and magnetic field strengths are conjectured to be conserved, then $E= -\nabla \phi$, where $\phi=-E\, r$, and
\bea\label{eom2}
\mathbf{v}_f^2 - \mathbf{v}_i^2 &=& -\frac{2}{m} q \left[\phi(r_f) - \phi(r_i)\right].
\eea
In Eq (\ref{eom2}), we use $d\phi/dr=\nabla\phi dr$. Therefore, the total energy that the charged particle gains through such an electric interaction reads 
\bea \label{energ22}
\epsilon &=& 2 q \left[\mathbf{r}_f\cdot \mathbf{E}_f - \left(\mathbf{r}_i\cdot\mathbf{E}_i\right)\frac{\mathbf{v}_i}{\mathbf{v}_f}\right] + m\, \mathbf{r}_i \cdot \mathbf{\dot{v}}_i \left(\frac{\mathbf{v}_i}{\mathbf{v}_f}\right),
\eea
Suppose that $E_i=E_f=0$, i.e. no field acceleration would take place, then $v_i=v_f$ and the total energy is simply given by the initial value $m \dot{v}_i r_i$. The subscripts $i$ and $f$ stand for initial and final, respectively. \\

To assure that the total acceleration does not dependent on the gyroradius, Eq. (\ref{lorentz1}) is expanded around $\mathbf{r}_L=0$ and then temporarily averaged over the whole gyration period of time,
\bea\label{eq:vparallel1}
\frac{d}{d\,t}\mathbf{v}_{\parallel} &=& \frac{q}{m}\mathbf{E}_{\parallel} - \frac{\mu}{m}\frac{d}{d\,r}B + \mathbf{v}_L \frac{d}{d\,t}\hat{\mathbf{e}}_L,
\eea
where $\hat{\mathbf{e}}_L$ is the unit vector along $\mathbf{v}_L$. Under certain assumptions, last term in both Eqs. (\ref{lorentz1}) and (\ref{eq:vparallel1}) can be removed. In order to discuss the physical meaning of each term in Eq. (\ref{eq:vparallel1}), we use - hereafter - reduced guiding center equations. Then, $\mu=(m v^2_{\perp}/2)/B\equiv W_{\perp}/B$ is valid. The latter is known as ''first adiabatic'' invariance. It assures that the ratio of kinetic energy $W_{\perp}$ of the Larmor motion (gyromotion) to the gyrofrequenvy is conserved. $\mu$ is proportional to the current due to Larmor motion through the relation $\mu=I A= (q\Omega_c/2\pi)(\pi r_L^2)$. Therefore, $\mu$ gives the magnetic moment.

Assuming that the magnetic field is static and the charged particle moves on a frame, in which the perpendicular velocity forms a cyclotron (helix or spiral) motion or Larmor motion, then dotting Lorentz equation, Eq. (\ref{lorentz1}),  by $\vec{v}$, and omitting the second term lead to 
\bea
\frac{d }{d t} m v_L^2 = 2 q \mathbf{v}_{\perp} \cdot \mathbf{E}_{\perp} + mv_L^2 \frac{\mathbf{v}_{\parallel}}{B}\frac{\partial B}{\partial r},   
\eea
where $\mathbf{v}_{\perp}$ is the vertical Larmor orbital velocity. Such a velocity is orthogonal to the $\mathbf{B}$--field, then the multiplying the following equation of motion \cite{larmor},    
\bea
m \frac{dv_{\parallel}}{dt} &=& q E_{\parallel} - \frac{1}{2} m v_{L}^2 \frac{1}{B} \frac{\partial B}{\partial r} = -q \frac{\partial \phi}{\partial r} - \mu \frac{\partial B}{\partial r}, 
\eea
by $v_{\parallel}$ results in conservation of the total energy of the gyration center
\bea\label{eq:timeDer1}
\frac{d}{dt} \left[ \frac{1}{2}\, m\, v_{\parallel}^2 + q\, \phi + \mu\, B\,\right]&=&0.
\eea
Suppose that the charged particle experiences an increasing magnetic field. As discussed above, $\mu$ is conserved. Then, $W$ must be an increasing quantity. It results in an increasing $\mathbf{v}_{\perp}$--components, as well. Since $\mathbf{B}$ can not add any additional work, then $\mathbf{v}_{\parallel}$ must increase in order to conserve the total kinetic energy. The particle is trapped inside this field, i.e. a magnetic mirror, as long as $\mathbf{v}_{\parallel}$ is sufficiently low. Otherwise, it experiences a force in the direction away from $\mathbf{B}$. Once again, the formalism is given in the non--relativistic limit. This would not prevent or exclude the relativistic one.

\subsection{Fermi mechanisms plus field interactions}

We assume that the cosmic charged particles speeds through astrophysical surroundings or conditions which are offering them acceleration via electromagnetic fields. Afterward, the particles are assumed to enter through regions of magnetic irregularities. According to the second-order Fermi mechanism, the magnetic irregularities result is elastic scattering.  First, we take into consideration cosmic charged particles which are accelerated from rest, i.e. their velocities increase from $0$ to $v$, Eq.~(\ref{eq:ebv}). In this case, the ansatz $\mathbf{v}\rightarrow \mathbf{v}\pm 2\mathbf{u}$ can be applied in the second--order Fermi mechanism, where both $\mathbf{v}$ and $\mathbf{u}$ are obviously aligned. Then
\bea\label{eq:eF2}
\epsilon_F &=& \sqrt{\sqrt{2} \, m\,r \, q \left(\mathbf{E}_{\parallel} - \mathbf{V}\times \mathbf{B}\right)} \; (v\pm 2u),
\eea
where $u$ is the velocity of frame of reference of the magnetic cloud. Furthermore, under the assumption that the cosmic charged particle diffuses upstream or downstream (first-order Fermi mechanism) while preserving its energy in rest frame, the particle's initial energy will also be preserved proportional to $\beta=v/u$ ratio, i.e, $v$ in Eq. (\ref{eq:eF2}) is to be replaced by $\beta\,u$. 

Then, we take into consideration the other case that the cosmic charged particle enters the magnetic irregularities and/or interacts with the electromagnetic fields, where its initial velocity $v_i$ is finite. 
\bea\label{eq:feildFermi1}
\epsilon_F &=& \frac{\mathbf{v}_i}{\mathbf{v}_f\pm 2\mathbf{u}} \left[m \, \mathbf{r}_i \cdot \mathbf{\dot{v}}_i  - 2 q \mathbf{r}_i \cdot \mathbf{E}_i \right] + 2 q \mathbf{r}_i \cdot \mathbf{E}_f
\eea

In the previous expressions, for example Eq. (\ref{energ12}) and (\ref{energ22}), we explicitly take into consideration cases in which the acceleration occurs due to the $\mathbf{E}$ component which is oriented parallel to $\mathbf{B}$, where both electromagnetic fields are uniform. It is essential to mention here that such a uniform electric field would not be impossible to be maintained in magnetic clouds or matter clumps. The existence of uniform magnetic fields is well--known. We could assume that the conductivity would not likely be high so that the cosmic charged particles would not disturb the cloud's overall charge neutrality. To be convinced about the existence of $\mathbf{E}_{\parallel}$, we mention just one example. In the Earth's ionosphere \cite{ionospahre}, $\mathbf{E}_{\parallel}$ has been registered, where the auroral particle acceleration is believed to be resulted in by this field.   \\

According to present model, the final energy that the cosmic charged particle would gain, is accumulated from different sources. Fermi mechanisms are obvious sources. In present work, plasma field interactions are suggested as {\it additional} sources. The latter dates back to eighties of last century, when Hillas introduced his well--known graph \cite{hillas1} and a very special configuration that magnetic and electric forces are perpendicular to each other has been studied. This apparently would limit the gyration to a concrete minimum value.
In present work, a general configuration for both electromagnetic fields is taken into consideration. Also, it is conjectured that various interaction types might be possible in one astrophysical object. In the section that follows, we introduce examples on drift forces depending on the nonuniformality and nonhomogeneity of both electromagnetic fields. Practically, such drifts would be substituted in Eqs. (\ref{eq:eF2}) and (\ref{eq:feildFermi1}).

\section{Examples of electromagnetic drifts}

Equation (\ref{eq:ve}) gives a general expression for the drift velocity. 
To calculate other drift velocities of the gyration center due to the components of $\mathbf{E}$ whose direction is perpendicular to $\mathbf{B}$, Eq. (\ref{eq:timeDer1}) can be utilized. The latter apparently assumes that $\mathbf{E}$ (or eventually any constant $\mathbf{F}$) and $\mathbf{B}$, as given in Eq. (\ref{lorentz1}), can be spatiotemporal dependent. In following subsections, we elaborate various spatiotemporal dependencies. The resulting drift forces are to be inserted into Eq. (\ref{eq:eF2}), in order to calculate the final energy, that the cosmic charged particles would gain through Fermi and field accelerations.

\subsection{Spatially varying magnetic field strength $\mathbf{B}$ }

For a weak spatially varying magnetic field $\mathbf{B}$, we assume that $\mathbf{B}$ has components in $\mathbf{z}$-direction. For simplicity, we assume that $\mathbf{B}$ varies in $\mathbf{y}$-direction, only. Then, the gradient with respect to $\mathbf{y}$ implies that
$\mathbf{B}_z >> \mathbf{r}_L \, d\, \mathbf{B}_z/d\, y$, 
which means that the field strength would likely have a Taylor expansion around $y=0$ over region or distance $\mathbf{y}\leq\mathbf{r}_L$. This leads to 
$\mathbf{B}_z(\mathbf{y})=B_0+\mathbf{y}\, d\mathbf{B}_z/dy+\cdots$.
Under the influence of this large and long-ranged magnetic field, the curvature of the particle's circular orbit becomes tighter so that it can be transferred into cycloid. The negative gradient is directed from stronger to weaker magnetic field regions, i.e., Larmor radius increases. According to this motion, the drift force is perpendicular to $\mathbf{B}$ and $\nabla |\mathbf{B}|$. Since we assume that the electric field vanishes, then $d\mathbf{v}/dt=(q/m) \mathbf{v} \times \mathbf{B}$, $x$-component averaged over the gyromotion vanishes as well and averaged $x$-component equals 
\bea
\left\langle \frac{d\,v_y}{d\,t}\right\rangle &=&\pm\, q\,\frac{\mathbf{v}_{\perp}\, \mathbf{r}_L}{2m} \frac{d\mathbf{B}_z}{dy}. 
\eea
As given above, we assume that the force $\mathbf{F}$ makes a $\pi/4$ angle with the acceleration. Then $<\cos^2(\Omega_c t)>=1/2$ while $<\cos(\Omega_c t)>=0$.  
Therefore, the ''grad $\mathbf{B}$'' drift velocity in three dimensions (Jackson's result) reads
\bea
\mathbf{v}_{\nabla \mathbf{B}} &=&\pm \frac{\Omega\, \mathbf{r}^2_L}{2} \left(-\frac{\nabla |\mathbf{B}| \times \mathbf{B} }{B^2}\right),
\eea
which can be seen as a result from the dipole force in $\mathbf{B}$ gradient. It is associated with a change in the magnetic field strength $\mathbf{B}$. The direction is perpendicular to both $\nabla |\mathbf{B}|$ and $\mathbf{B}$. The Ampere's law for constant or vanishing $\mathbf{E}$ reads $\nabla \times \mathbf{B}=\mu_0\mathbf{J}$. Apparently, this law seems to describe the current that generates the magnetic field. Finally, we get the ''grad $\mathbf{B}$'' drift force 
\bea \label{force-nablaB}
\mathbf{F}_{\nabla \mathbf{B}} &=& \pm\, q\,\frac{\mathbf{v}_{\perp}\, \mathbf{r}_L}{2} \frac{d\mathbf{B}_z}{dy},
\eea

\subsection{Non-uniform $\mathbf{B}$ and vanishing $\mathbf{E}$}

There is one further drift associated with non-uniform $\mathbf{B}$ and vanishing $\mathbf{E}$ fields. The charged particle moving along a curved magnetic field line is conjectured to experience a centripetal acceleration. To study this problem, two-dimensional cylindrical coordinate system (${\cal R}$,$\Phi$) is defined. The origin of the cylindrical coordinate system is located at the field line center of curvature and lying on the plane of the field lines. The radial location of the chosen point in this system is identical with the radius of local curvature of the field lines ${\cal R}$. Then $\hat{\Phi}=\hat{B}$ and $\hat{B}\cdot\nabla\hat{B}=\hat{\Phi}\cdot\nabla\hat{\Phi}=-\hat{{\cal R}}/{\cal R}$. The ''curvature'' drift velocity in three dimensions is
\bea
\mathbf{v}_c &=& \frac{1}{q\, B^2} \left(m\, \mathbf{v}^2_{\parallel gc} \frac{\hat{{\cal R}}}{{\cal R}}\right) \times \mathbf{B},
\eea
which is directed either into or out of the page depending the sign of the electric charge $q$. The centripetal force reads 
\bea \label{force-cp}
\mathbf{F}_{cp} &=& m\, \mathbf{v}^2_{\parallel gc} \frac{\hat{{\cal R}}}{{\cal R}}.
\eea

\subsection{Time-varying electric field strength $\mathbf{E}$}

For slightly time-varying electric field $\mathbf{E}$, the electric and magnetic field strength would read 
\bea
\mathbf{E}(\mathbf{x},t) &=& E_0(\mathbf{x})\;\exp(i\Omega t), \\
\mathbf{B}(\mathbf{z}) &=& B\, \mathbf{z}.
\eea
We notice that equation (\ref{eq:timeDer1}) would result in an equation of motion as follows. 
\bea
\frac{d^2 \mathbf{v}_{\perp}}{d\, t^2} &=& -\Omega_c^2 \left[\mathbf{v}_{\perp} - \frac{q}{|q|} \frac{i\Omega}{\Omega_c} \frac{\mathbf{E}_{\perp}(\mathbf{x},t)}{\mathbf{B}(\mathbf{z})}\right] \equiv -\Omega_c^2 \left[\mathbf{v}_{\perp} - \mathbf{\tilde{v}}_{\parallel}\right],
\eea
which obviously describes a harmonic motion. With the assumption that $\mathbf{E}$ is slight varying in time, the ''polarization'' drift velocity takes the form
\bea
\mathbf{v}_p &=& \frac{m}{q B^2}  \frac{d E_{\perp}(\mathbf{x},t)}{d\, t} = -\frac{1}{\Omega_c \, B} \left(\frac{d}{d\,t} \mathbf{V}\right) \times \mathbf{B},
\eea
where $\mathbf{V}$ is the generic drift velocity and the direction is defined by the sign of the electric charge $q$, as given in Eq. (\ref{rrl_special}). 
This drift motion obviously differs from other drifts. It even can inhere in it other drift motions. It is not allowed to continue indefinitely. Examples for temporarily varying electric field $\mathbf{E}$ are not rare. For example an oscillatory $\mathbf{E}$, which would result in $\mathbf{v}_p$. The latter oscillates normally out of phase. Because of the $m$-dependence, this drift is also called ''inertia'' and can be neglected for light particles. Then, as given above, the ''polarization'' drift force reads
\bea \label{force-c}
\mathbf{F}_p &=& -\Omega^2\, m\, \left[\mathbf{r}_{\perp}-\tilde{\mathbf{r}}_{\parallel}\right].
\eea

\subsection{Time-varying magnetic field strength $\mathbf{B}$}

According to Faraday's law, the time-varying magnetic field $\mathbf{B}$ generates an electric field $\mathbf{E}$, $d\,\mathbf{B}/dt=-\nabla\times\mathbf{E}$. Considering a vector $\mathbf{l}$ along the perpendicular trajectory, then the time evolution of the particle's kinetic energy reads
\bea
\frac{1}{2} m \frac{d}{d\,t} \mathbf{v}_{\perp}^2 &=& q \mathbf{E}\cdot \frac{d\, \mathbf{l}}{d\,t},
\eea
which leads to change the initial kinetic energy to
\bea
\frac{1}{2} m \left( \mathbf{v}_{\perp,f}^2-\mathbf{v}_{\perp,i}^2 \right) &=& 2\pi\,\frac{\mu}{\Omega_c} \frac{d\, \mathbf{B}}{d\,t}.
\eea 
The r.h.s. gives the energy added to the initial kinetic energy by time-varying magnetic field $\mathbf{B}$. The ''grad $\mathbf{E}$'' drift force reads
\bea \label{force-nablaE}
\delta \mathbf{F}_{\nabla\mathbf{E}} &=& 2\pi\,\frac{\mu}{\Omega_c\,\mathbf{r}} \frac{d\, \mathbf{B}}{d\,t}.
\eea

\subsection{Generic drifts}

One can think of two other cases, in which both fields are either spatially or temporarily varying. That one is spatially or temporarily varying while the other is the opposite would give two additional cases. 
Various drift motions are possible under action of the generic force $\mathbf{F}=\mathbf{F}_{\parallel}\cos(\theta)+\mathbf{F}_{\perp}\sin(\theta)$. The component $\mathbf{F}_{\parallel}\cos(\theta)$ seems to add to the acceleration along $\mathbf{B}$ and the component $\mathbf{F}_{\perp}\sin(\theta)$ add to the drift motion velocity. This leads to the conclusion that the final acceleration under the influence of the field interactions is composed of three parts:  
\begin{itemize} 
\item direct acceleration under the influence of $\mathbf{F}_{\parallel}\cos(\theta)$,
\item drift acceleration by $\mathbf{F}_{\perp}\sin(\theta)$ and
\item gyration by $\mathbf{B}$ itself. 
\end{itemize}
In Eq. (\ref{eq:feildFermi1}), $\mathbf{E}$ can be replaced by the summation of the forces given in Eqs. (\ref{force-nablaB}), (\ref{force-cp}), (\ref{force-c}) and (\ref{force-nablaE}). The various hypothetical astrophysical sources would sanctify to conclude that the cosmic particles accumulatively gain their final energy through the participation in a wide range of various interactions. 


\section{Discussion}

The Fermi mechanism \cite{fermi49} is to be illustrated as the cosmic charged particles are accelerated through random elastic collisions with the irregularities of the interstellar objects, such as magnetic clouds and matter clumps, etc. Therefore, the sources for the ultra high energy are likely time-dependent magnetic and induced electric fields. The energy that the cosmic charged particles would gain depends only on $\mathbf{u}$, the velocity of the magnetic clouds, but not on the magnetic field strnegth $\mathbf{B}$ and/or electric charge $q$. Because of the rare elastic collisions, this mechanism is criticized to be too slow to attain UHECR. The other concern is going on the effects of Coulomb interactions on the energy loss of the cosmic charged particles injected at relatively low energy. The Coulomb interactions use a large portion of this gained energy. Astrophysical objects with turbulent magnetic fields would offer a realization of matter clumps, that homogeneously distributed in the galactic space and move with the Alfven wave velocity $v_A$. The shock scatterings through the expansion of matter matter clumps flowing at speeds larger than the speed of sound, like remnant supernovae, are likely frequent. The acceleration according to this mechanism is much more efficient than the stochastic one.

So far, we summarize that the electromagnetic drift motions would be possible sources to accelerate cosmic charged particles. Since various types of field interactions can also exist in cold plasmas, non--relativistic formalism for different the types of electromagnetic interactions is given. The translation to a relativistic version is straightforward. The elastic and stochastic scatterings, the Fermi acceleration mechanisms, play their essential role in this model. Furthermore, general configurations for both electromagnetic fields are taken into consideration. It has been conjectured that various interaction types might be possible in one astrophysical object. Examples on drift forces depending on the nonuniformality and nonhomogeneity of both electromagnetic fields have been discussed.  The final energy, that the cosmic charged particle would gain, is assumed to stem from different sources. Fermi mechanisms are obvious sources. Plasma field interactions are suggested as {\it additional} sources. Obviously, the latter date back to eighties of last century, when Hillas introduced his well--known graph \cite{hillas1} and a very special configuration that magnetic and electric forces are perpendicular to each other has been studied. This apparently would limit the gyration to a concrete minimum value. 

The present model is applied on different astrophysical objects, in order to calculate the energy gained by a singly charged and massive particle \cite{clacs}. It has been found that the order of the electromagnetic field acceleration is zero--order, while Fermi mechanisms apparently describe first-- and second--order relative velocities \cite{orders}.



\begin{thebibliography}{99}
\bibitem{cr_comps} The most dominant components of ultra high energy cosmic rays are the protons.
\end{thebibliography}
\end{document}